\documentclass[twocolumn]{pasj00}
\draft
\Received{$\langle$reception date$\rangle$}
\Accepted{$\langle$acception date$\rangle$}
\Published{$\langle$publication date$\rangle$}
\SetRunningHead{Izawa et al.}{Suzaku observation of HESS~J1536$-$645}

\usepackage{times}
\usepackage{threeparttable}
\usepackage{subfigure}
\begin{document}

\title{Suzaku observations of the old pulsar wind nebula
candidate HESS~J1356$-$645}
\author{Masaharu \textsc{Izawa}$^{1,2}$,
Tadayasu \textsc{Dotani}$^{2,1,3}$,
Takahisa \textsc{Fujinaga}$^4$,
Aya \textsc{Bamba}$^5$,
Masanobu \textsc{Ozaki}$^2$,
and Junko S.\ \textsc{Hiraga}$^6$
}
\affil{$^1$
Department of Physics, Tokyo Institute of Technology,
2-12-1 Ookayama, Meguro-ku, Tokyo 152-8550, Japan}
\affil{$^2$
Institute of Space and Astronomical Science/JAXA,
3-1-1 Yoshinodai Chuo-ku, Sagamihara, Kanagawa 252-5210, Japan}
\affil{$^3$
Space and Astronautical Science, School of Physical Sciences, SOKENDAI,
3-1-1 Yoshinodai Chuo-ku, Sagamihara, Kanagawa 252-5210, Japan}
\affil{$^4$
Tsudakoma Corp., 5-18-18 Nomachi, Kanazawa 921-8650, Japan}
\affil{$^5$
Department of Physics and Mathematics, Aoyama Gakuin University,
5-10-1 Fuchinobe Chuo-ku, Sagamihara, Kanagawa, 252-5258, Japan}
\affil{$^6$
The University of Tokyo, 7-3-1 Hongo, Bunkyo-ku, Tokyo 113-0033, Japan}
\email{dotani@astro.isas.jaxa.jp}

\maketitle

\begin{abstract}
A largely extended X-ray emission was discovered around the pulsar PSR~J1357$-$6429 
with the Suzaku deep observations.
The pulsar, whose characteristic age is 7.3~kyr, is located within 
the TeV $\gamma$-ray source HESS J1356$-$645.
The extended emission is found to have a 1$\sigma$ X-ray size of
$\sim$4~arcmin, or $\sim$3~pc 
at 2.4~kpc, with a small offset from the pulsar.
Its X-ray spectrum is well reproduced by a simple power-law model
with a photon index of $1.70_{-0.06}^{+0.07}$.
No significant spatial variation was found for the X-ray photon index as a function of distance from the pulsar.
%No significant spatial variation was found in the photon indices according to the distance from the pulsar.
We conclude that the extended emission is associated to the
pulsar wind nebula of PSR~J1357$-$6429.
This is a new sample of largely extended nebulae around middle-aged pulsars.
We discuss the evolution of this PWN according to the relic PWN scenario.
\end{abstract}

\section{Introduction}

Pulsar wind nebulae (PWNe) consist of ultra-relativistic electrons and positrons
accelerated by the pulsar termination shocks.
Young PWNe are bright X-ray emitter via synchrotron process 
and are well studied in X-rays
\citep[for example]{2006ARA&A..44...17G,2008AIPC..983..171K,2010ApJ...709..507B}.
However, they become fainter in X-rays when the central pulsars get older 
\citep{2009ApJ...694...12M}.
Only limited studies exist for the older sources.

The atmospheric Cherenkov telescope
for TeV (10$^{12}$ eV) gamma-rays, H.E.S.S. (High Energy Stereoscopic System),
conducted Galactic plane survey since 2004
covering effectively the whole inner Galaxy,
and discovered about fifty new sources in the very-high-energy (VHE)
$\gamma$-ray band
\citep{2009arXiv0907.0768C}.
Because PWNe occupy the largest population of the identified Galactic 
VHE sources so far, majority of the new sources may also be the PWNe
\citep{2006ApJ...636..777A}.
A flow of high energy electrons and positrons in the PWNe contributes
VHE $\gamma$-rays emission through inverse Compton scattering
and X-rays through synchrotron emission.  
It became possible to estimate physical parameters of the PWNe, eg.\ magnetic
field, by combining the VHE $\gamma$-ray and X-ray data.
Thus the multi-wavelength studies of PWNe in VHE $\gamma$-rays and X-rays are
very important to reveal their nature.

Recent studies of PWNe revealed the presence of extended X-ray emission
with low surface brightness around the pulsar.
Such emission may be produced by the ultra-relativistic electrons\footnote{Hereafter, we mention
only electrons but should be interpreted as electrons and positrons.}
accelerated in the termination shocks.
If so, its extension should correspond to the electron diffusion length during the 
energy-loss time scale.
The energy-loss time scales of the X-ray emitting electrons may be 2--10~kyr    % 2-12 kyr was changed to 2-10 kyr (T.D.)
assuming the magnetic field of the Galactic plane, 3--10~$\mu$G.
This means that the electrons can travel $\sim$2~pc at the most.
However, Suzaku observations showed that
some PWNe had the extended emission with the size of more than 10~pc
\citep{2010PASJ...62..179A,2009PASJ...61S.189U}.
Furthermore, \citet{2010ApJ...719L.116B} reported that
the size of the extended emission from PWNe kept expanding
up to $\sim$100~kyr in their characteristic ages.
However, it is poorly understood
how the emission region is formed and how it expands.

One of the key parameters to understand how the extend emission region is
formed and evolves is the energy loss rate of the electrons.
Because the ultra-relativistic electrons lose energy during diffusion, 
the non-thermal emission from the PWNe, which is well represented
by a power-law, is expected to become softer with the distance from the pulsar.
But this is sometimes not the case.
Some PWNe showed no systematic variations in the photon indices of the 
power-law component with the distance from the pulsar
\citep{2010PASJ...62..179A,2009PASJ...61S.189U}.
This result may be explained if the ultra-relativistic electrons can 
diffuse with little loss of energy.
However, it is not clear at present whether such lossless diffusion is plausible or not.

The pulsar PSR~J1357$-$6429
($P=166$ ms, $\dot{E}=3.1\times10^{36}$~erg~s$^{-1}$,
surface magnetic field $B=7.8\times10^{12}$ G)
was discovered in the Parkes Multibeam Pulsar Survey
\citep{2004ApJ...611L..25C}.
The characteristic age, $\tau$,  is 7.3~kyr.
%indicates that the pulsar is one of the youngest pulsars known.
The distance of 2.4~kpc was estimated from the pulsar's dispersion measure
(DM $=127.2$ cm$^{-3}$ pc) and the Galactic free electron density model
\citep{2002astro.ph..7156C}.    
PSR J1357$-$6429 is located within the TeV $\gamma$-ray source
HESS J1356$-$645 \citep{2011A&A...533A.103H}.
The extended TeV emission was detected up to \timeform{12'}
from the peak of the TeV surface brightness,
located at R.A. = \timeform{13h56m}, dec.\ = \timeform{-64D30'} (J2000).
The TeV source was most likely powered by the pulsar
because the separation of the center of the TeV source from the pulsar is 
only by $\sim$\timeform{7'}
and the TeV luminosity is $\sim6\times10^{33}$~erg~s$^{-1}~\sim0.002\dot{E}$.      
At radio frequencies, \citet{1997MNRAS.287..722D} reported
the extended emission from the supernova remnant (SNR) candidate G309.8$-$2.6,
which is spatially coincident with HESS J1356$-$645.
XMM-Newton and Chandra observed HESS J1356$-$645 in 2005 August for15~ks
and in 2005 November for 33~ks, respectively.
\citet{2007nscs.confE...2Z} and \citet{2007A&A...467L..45E} reported that
the pulsar spectrum likely consists of thermal and non-thermal components.
\citet{2007nscs.confE...2Z} also reported a tail-like extended emission
in the HRC-S image.
\citet{2012ApJ...744...81C} suggested that
this structure could be a pulsar jet or a part of the pulsar tail.  
\citet{2012A&A...540A..28D} found a possible optical counterpart of this pulsar
and estimated the proper motion velocity to be $\sim$2000~km~s$^{-1}$
to the north-eastern direction.

The X-ray Imaging Spectrometer (XIS; \cite{2007PASJ...59S..23K})
on board the Suzaku satellite \citep{2007PASJ...59S...1M}
has a low and stable background
in contradiction to the limited sensitivity of XMM-Newton and Chandra
for diffuse sources.
Actually, \citet{2009PASJ...61S.189U} and \citet{2013PASJ...65...61F}
reported that the sizes of the extended emission around VHE gamma-ray sources
determined by Suzaku were larger than
the ones determined by XMM-Newton or Chandra.
Therefore, XIS is ideal for observing faint and extended X-ray emission
around HESS J1356$-$645.
In addition, because it is located away from the Galactic plane,
the observation would be less influenced
by the Galactic Ridge X-ray Emission (GRXE)\@.
Thus, Suzaku observations would show new morphology of HESS J1356$-$645.

In this paper, we show the first Suzaku result on HESS~J1356$-$645.
In \S\ref{sec:obslog},
we present the observation log and data reduction.
Imaging and spectral analyses are shown in \S\ref{sec:results}.
We discuss the origin of the diffuse emission
around PSR~J1357$-$6429 in \S\ref{sec:discussion}.
Errors indicate single parameter 90\% confidence regions
throughout the paper.

\section{Observations and Data Reduction}
\label{sec:obslog}
We observed the region around HESS J1356$-$645
with Suzaku in 2013 January and February.
The observations were carried out
with the two pointings in the east and west of HESS J1356$-$645
to cover the extended X-ray emission.
The total exposures were 55.7~ks and 51.2~ks for the east and west region,
respectively.
Table \ref{tab:obslog} lists the journal of the Suzaku observations. 

The Suzaku observations were performed
with XIS in 0.3$-$10~keV and Hard X-ray Detector
(HXD; \cite{2007PASJ...59S..35T}; \cite{2007PASJ...59S..53K})
in 13$-$600~keV\@.
XIS is located at the focal plane of the X-Ray Telescopes
(XRT; \cite{2007PASJ...59S...9S}),
which provides a spatial resolution of $\sim\timeform{2'}$
in a half-power diameter (HPD)
and a field-of-view (FOV) of $\timeform{18'} \times \timeform{18'}$.
The XIS system consists of one back-illuminated (BI) CCD camera (XIS1)
and three front-illuminated (FI) CCDs (XIS0, 2, and 3).
One of the front-illuminated CCDs, XIS2, was not usable
at the time of our observations
since it suffered from fatal damage on 2006 November 9.
The XIS instruments were operated in the normal full-frame clocking mode
(a frame time of 8~s)
with Spaced-row Charge Injection \citep{2008PASJ...60S...1N}.
The HXD consists of silicon PIN photodiodes capable of observations
in the 13$-$70~keV band
and GSO crystal scintillators which cover the 40$-$600~keV band.
Because we could not find any pulsations in the HXD data
with good timing resolution of 61~$\mu$s \citep{2008PASJ...60S..25T},
we focused on the XIS data analysis in this paper.

We used data sets processed 
by a set of software of the Suzaku data processing pipeline 
(version 2.5.16.29).
The telemetry saturated time was excluded in the pipeline processing.
Basic analysis was done using the HEASOFT software package (version 6.12).
We used cleaned event files, in which standard screening was applied.
The standard screening procedures include event grade selections,
and removal of time periods such as spacecraft passage of
the South Atlantic Anomaly, intervals of small geomagnetic cutoff rigidity,
and those of a low elevation angle from the earth limb.
Specifically, for the XIS, elevation angles larger than 5$^{\circ}$
above the Earth and larger than 20$^{\circ}$
from the sunlit Earth limb are selected. 

\begin{longtable}{*{4}{c}}
\caption{Journal of the observations}
\label{tab:obslog}
\hline
\hline
Obs ID & Observation date & Aim point\footnotemark[$*$] & Effective exposure (ks) \\
\hline
507019010 & 2013/01/26$-$2013/01/28 & (\timeform{309.D99}, \timeform{-2.D55}) & 55.7  \\
507020010 & 2013/02/17$-$2013/02/18 & (\timeform{309.D73}, \timeform{-2.D48}) & 51.3  \\
\hline
\multicolumn{4}{l}{\footnotemark[$*$] In Galactic coordinates.} \\
\endlastfoot
\end{longtable}

\section{Analysis and Results}
\label{sec:results}

\subsection{Background spectrum}
\label{sec:bgspectrum}
\begin{figure}[ht]
  \begin{center}
\includegraphics[width=80mm]{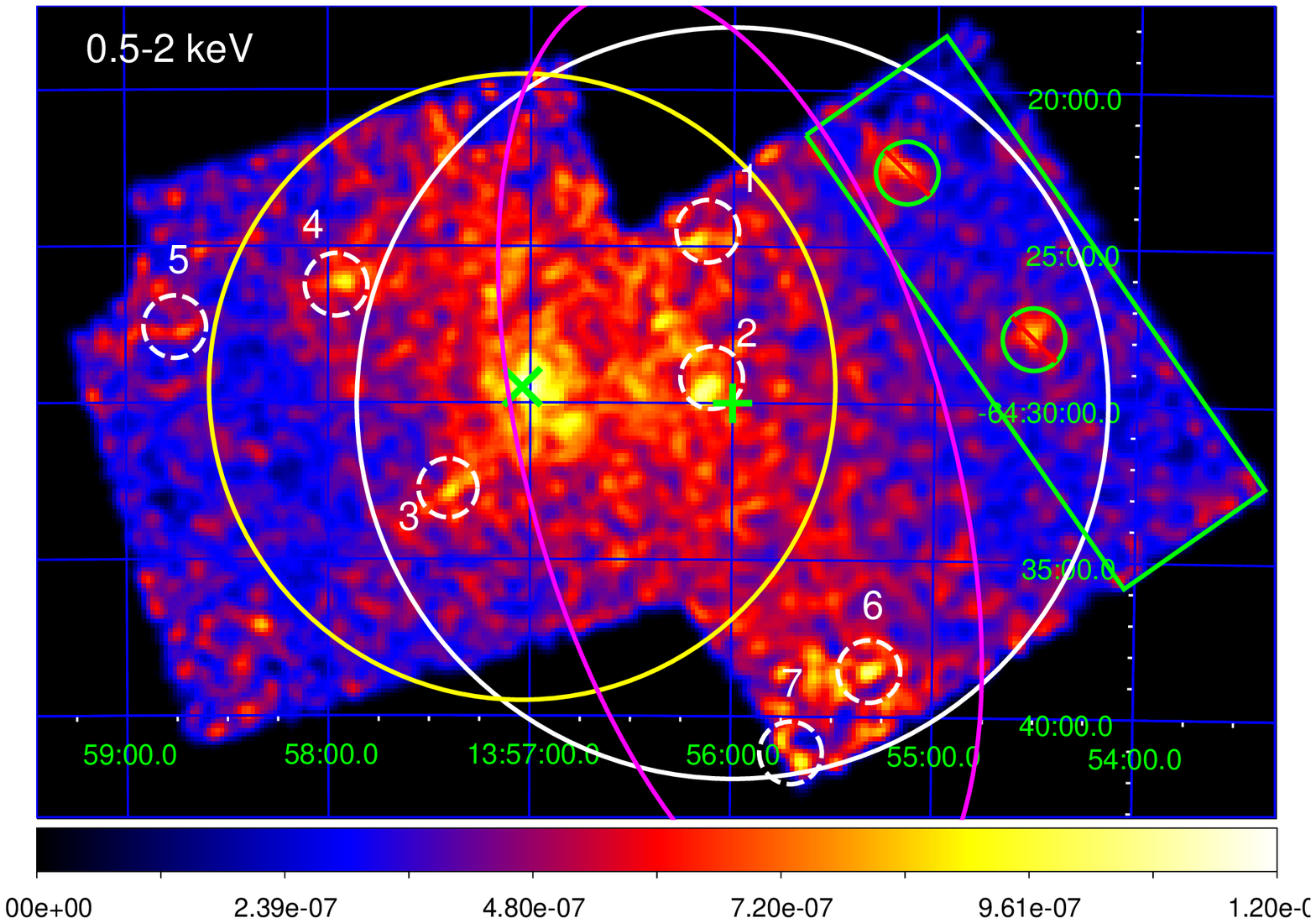}
\includegraphics[width=80mm]{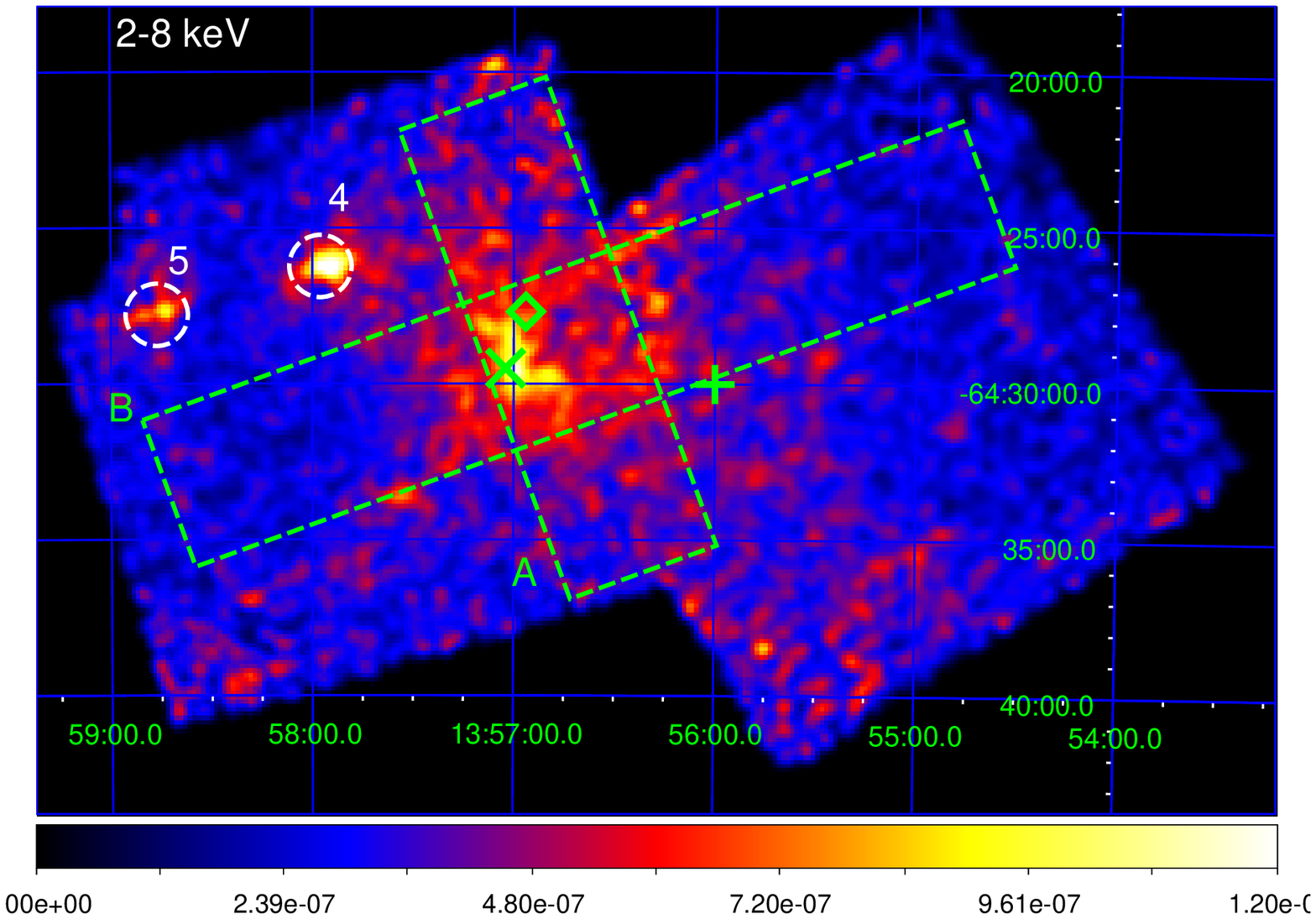}
 \end{center}
  \caption{%
XIS image of the HESS~J1356$-$645 field in the 0.5$-$2~keV (top)
and the 2$-$8~keV (bottom) bands, respectively.
These images were smoothed using a Gaussian function of $\sigma=2$~arcmin.
The color scale is in units of counts~s$^{-1}$/69.5~arcsec$^{2}$.
A corner illuminated by the calibration source was removed.
The peak position of the VHE $\gamma$-ray emission and PSR J1357$-$6429 is
shown with $+$ and $\times$, respectively.
The white and yellow circle (0.5$-$2 keV) indicates
the extent of the VHE source (\timeform{12'} in radius)
and
%the source region of spectra
the region used for the spectral extraction
(\timeform{10'} in radius; see \S3.3.1.),
respectively.
Note that the areas off the chip were not used in the analysis.
The magenta ellipse marks the extent of the radio SNR candidate G309.8$-$2.6
(Parkes 2.4~GHz; \cite{1995MNRAS.277...36D}).
The background region is shown with the green solid rectangle
(\timeform{5'.5}$\times$\timeform{18'}).
The sources 1$-$7 are catalogued by XMM-Newton.   
The diamond in the right panel indicate the peak position of the extended
emission  (see \S\ref{sec:extended_emission}).  
The rectangles in the right panel are the regions for the profile analysis.
}
  \label{fig:xisimage}
\end{figure}

Figure~\ref{fig:xisimage} shows XIS images of the region
around HESS J1356$-$645 in the 0.5$-$2 keV and 2$-$8 kev bands.
The Non X-ray background (NXB) generated
by \texttt{xisnxbgen} \citep{2008PASJ...60S..11T} was subtracted
from the images.
The images were binned by $8\times8$ pixels and smoothed
with a Gaussian function of $\sigma=2$~arcmin.
We corrected the vignetting effect by dividing the image with a flat sky image,
which was simulated using \texttt{xissim} \citep{2007PASJ...59S.113I}.
In this simulation, we assumed the input energy spectrum
as that extracted from the background region.
We explain details of this background spectrum below.   

\begin{figure}
  \begin{center}
    \includegraphics[width=80mm]{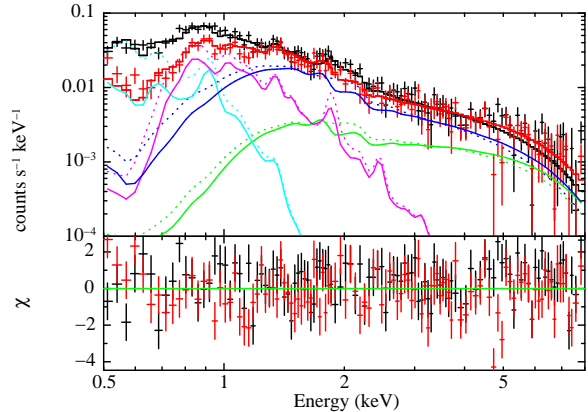}
  \end{center}
  \caption{%
Energy spectra of the background region are displayed with the best-fit model.
Red and Black crosses represent the FI and BI spectra, respectively.
The best-fit models for FI and BI are plotted with dotted and solid lines, respectively.
Individual components of the model are also shown for the thin thermal two-temperature plasma
(cyan and magenta),
a power-law (blue), and the CXB (green).
Vertical error bars of each data point represent the 1$\sigma$ error.
}
  \label{fig:bgdspec}
\end{figure}

The background region (as depicted by the green solid rectangle 
in figure~\ref{fig:xisimage}, which has an area of 91~arcmin$^2$)
was selected to minimize contamination from the diffuse emission.
Two point-like sources not catalogued by XMM-Newton
in the background region were removed
with \timeform{1'} radius circular regions.
After removing these point-like sources,
we extracted the background spectrum.
The background spectrum is shown in figure \ref{fig:bgdspec}.
There is no noticeable features of the emission lines
in the background spectrum.

We fitted the background spectrum
with the Cosmic X-ray Background (CXB) component
and a power-law component
for the remaining PWN emission.
For the latter component, we do not have to distinguish
whether
the extended emission is stray light of central bright PWN
or really extended and low surface brightness emission,
since it does not change our result,
as shown in section~\ref{sec:extended_emission}.
We assumed the CXB as a power-law of photon index 1.4
with the intensity $5.4\times10^{-15}$~erg~s$^{-1}$~cm$^{-2}$~arcmin$^{-2}$
in the 2$-$10 keV band \citep{1999ApJ...518..656U}.
The hydrogen column density of the CXB was fixed to
$0.84\times10^{22}$~cm$^{-2}$
determined by H\emissiontype{I} observations \citep{2005A&A...440..775K}.
Note that the CXB emission has $\sim 10$\% surface brightness fluctuation
\citep{cappelluti2012},
which is negligible compared with the background photons
(see Figure~\ref{fig:bgdspec}).
For the power-law component, the hydrogen column density, photon index,
and normalization were set to be free parameters.
The fitting was carried out for the FI and BI spectra simultaneously.
The detector responses (RMF) and telescope responses (ARF) were
generated by \texttt{xisrmfgen} and \texttt{xissimarfgen}
\citep{2007PASJ...59S.113I}, respectively. 

This model, however, was rejected
with $\chi^2_{\mathrm{\nu}}=1.6$ (d.o.f.=225)
with large residuals below 2 keV.
We then added a thin thermal plasma component
(APEC; \cite{2001ApJ...556L..91S}) to this model,
where metal abundances were fixed to the solar abundance
while the other parameters were set free.
The fit was improved to be $\chi^2_{\mathrm{\nu}}=1.4$ (d.o.f.=223),
but still residuals were seen in the lower energy band.
We then fitted the spectrum with two-temperature plasma
and a power-law plus the CXB.
The best-fit results are shown
in table \ref{tab:bgdspecresults} and figure \ref{fig:bgdspec}.    
The fit returned better $\chi^2_{\mathrm{\nu}}=1.3$ (d.o.f.=221),
with no significant residuals.
The low temperature component is 
consistent with the emission from the local hot bubble \citep{yoshino2009},
whereas the high temperature one with
the low-temperature component of the GRXE
\citep{ryu2009}.
Note that such modeling is
just for the reproduction of the background spectrum,
thus we do not discuss separate components further.

\begin{table}
\caption{Best-fit parameters to the background spectrum.}
\label{tab:bgdspecresults}
\begin{center}
\begin{tabular}{*{5}{c}}
\hline
\hline
 & apec1 & apec2 & power-law & CXB   \\ 
 \hline
$kT$\footnotemark[a] & $0.10^{+0.02}_{-0.01}$ & $0.58^{+0.07}_{-0.12}$ & -   & - \\ 
$\Gamma$\footnotemark[b] & - & - &  $2.2^{+0.1}_{-0.1}$ & (1.4) \\  
Flux\footnotemark[c] & 27 & 2.2 & 2.4 & (0.9) \\
\hline
N$_{\mathrm{H}}$\footnotemark[d] & \multicolumn{3}{c}{$0.53^{+0.10}_{-0.17}$} & 0.84 \\ 
\hline
$\chi^2_{\nu}$(d.o.f.) & \multicolumn{4}{c}{1.3(221)}  \\
\hline
\multicolumn{5}{l}{Note: Values in the parenthesis are fixed in the fitting.} \\
\multicolumn{5}{l}{\footnotemark[a]Plasma temperature [keV]} \\
\multicolumn{5}{l}{\footnotemark[b]Photon index } \\
\multicolumn{5}{l}{\footnotemark[c]Unabsorbed flux in the 0.5--10~keV band [$10^{-11}$~erg~cm$^{-2}$s$^{-1}$]}\\
\multicolumn{5}{l}{\footnotemark[d]Hydrogen column density [10$^{22}$cm$^{-2}$] } \\
\end{tabular}
\end{center}
\end{table}

\subsection{Image}
In both the soft and hard band images,
the position of the center of the brightest pixel is
R.A.$=$\timeform{13h57m01s.18}, decl$=$\timeform{-64D29'30''.4} (0.5$-$2~keV)
and R.A.$=$\timeform{13h56m59s.89}, decl$=$\timeform{-64D29'30''.4}
(2$-$8~keV), respectively.
The radio pulsar is located at R.A.$=$\timeform{13h57m02s.43} (2),
decl$=$\timeform{-64D29'30''.2} (1) \citep{2004ApJ...611L..25C}.
The difference of \timeform{8''} ($0.5-2$ keV) and \timeform{16''} ($2-8$ keV)
between the X-ray and radio positions is
smaller than the uncertainty in position of the XIS (\timeform{19''}
at the 90$\%$ confidence level, \cite{2008PASJ...60S..35U}).   
Thus we concluded that
the position of the brightest point-like source coincides with
that of the radio pulsar.

An extended emission is clearly seen around PSR J1357$-$6429.
Seven point-like sources in the soft band and two point-like sources
in the hard band are identified in the XMM-Newton serendipitous source
catalogue (3XMM).
These point-like sources are summarized in table \ref{tab:pointsource}.

\begin{longtable}{*{3}{c}}
\caption{Sources in the Suzaku fov from the XMM-Newton serendipitous source catalogue}
\label{tab:pointsource} 
\hline
\hline
Source number & Source~name & Source coordinate ($\mathrm{R.A._{J2000}}, \mathrm{Decl._{J2000}}$)    \\ 
\hline
1 & $\mathrm{3XMM~J135607.6-642430}$ & (209.0320,~--64.4086)   \\ 
2 &  $\mathrm{3XMM~J135704.4-642909}$ &(209.2685,~--64.4859)    \\ 
3 &  $\mathrm{3XMM~J135724.4-643244}$ &(209.3517,~--64.5456)    \\
4 &  $\mathrm{3XMM~J135757.2-642612}$ &(209.4887,~--64.4368)    \\
5 &  $\mathrm{3XMM~J135845.3-642734}$ &(209.6891,~--64.4596)    \\
6 &  $\mathrm{3XMM~J135518.7-643832}$ &(208.8282,~--64.6423)    \\
7 &  $\mathrm{3XMM~J135541.9-644109}$ &(208.9250,~--64.6859)    \\  
\hline
\endlastfoot
\end{longtable}

\subsubsection{Extended emission}
\label{sec:extended_emission}
\begin{figure*}
% \centering
\begin{center}
\includegraphics[width=80mm]{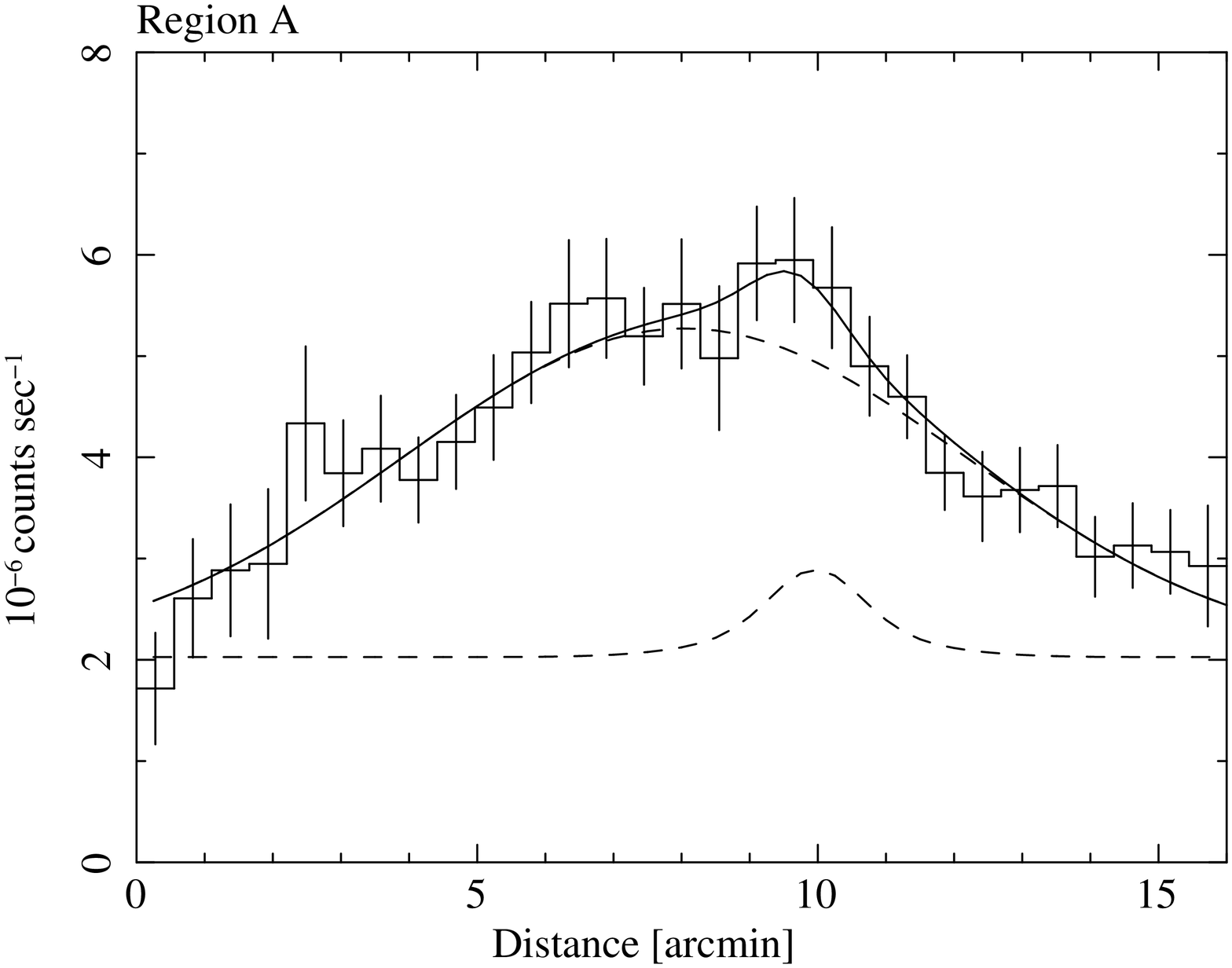}
\includegraphics[width=80mm]{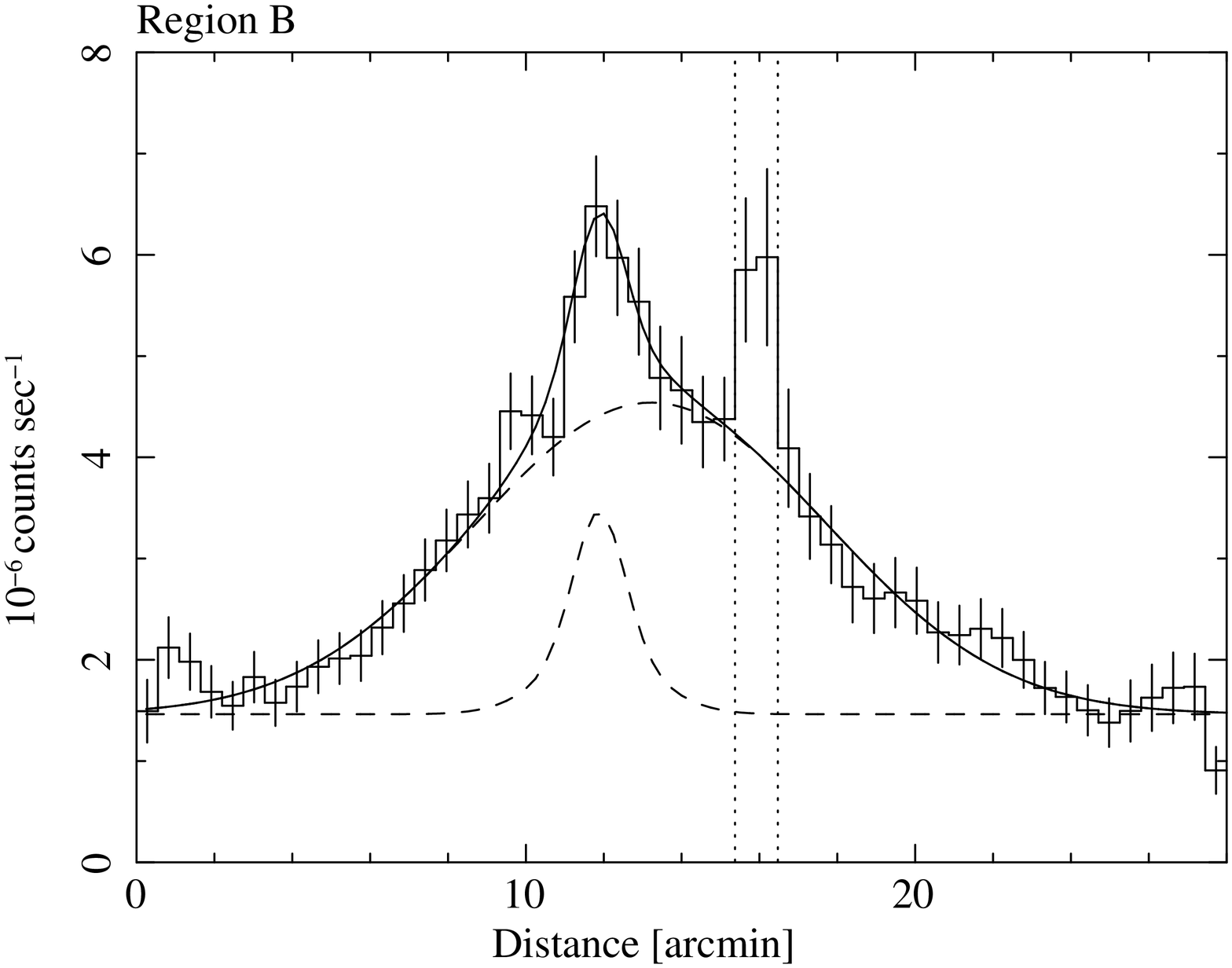}
\end{center}
  \caption{%
One-dimensional profiles of the diffuse emission.
The solid lines represent the best-fit model consisting of a Gaussian, PSF
and a constant function.
The PSF is approximated by a double Gaussian (see text for details).
The Gaussian and PSF are plotted as the broken lines.
The bright peak in the region B (data points subtended by the two vertical dotted lines) 
corresponding to a point source was ignored in the fitting.}
\label{fig:linirprofile}
\end{figure*}

In order to study the spatial distribution of the diffuse component,
we made 1-dimensional profiles in the 2$-$8~keV band
from the rectangular regions  shown 
in figure~\ref{fig:xisimage} with the broken line.
The regions were selected to be as wide as possible and to cross at right angles each other.
The size of the region is \timeform{16'} $\times$ \timeform{5'} (region A)
and \timeform{5'} $\times$ \timeform{28'} (region B), respectively.
From Figure~\ref{fig:xisimage}, one can see that
the pulsar is not at the center of the diffuse emission.
We set the center of the rectangular region not at the host pulsar
but at the rough center of the diffuse emission,
since our aim is to measure the extent of the diffuse emission.
Figure~\ref{fig:linirprofile} shows the 1-dimensional profiles of
region A and B.
We fitted the profiles with a model consisting of a Gaussian function (for the
diffuse emission), a point spread function (PSF; for the pulsar) 
and a constant component (for the background).
We approximated the PSF as a double Gaussian function,
whose parameters (each width = \timeform{1.3'}, \timeform{0.7'}
and the intensity ratio = 2.3) were determined
from the analysis of the 1-dimensional Suzaku image of the dwarf nova SS Cyg.
The same data of SS Cyg were also used for the flight calibration of the PSF
\citep{2009PASJ...61S..77I}.
So, the fitting model is given by
\begin{eqnarray}
y_{\mathrm{model}}=\mathrm{P_0}+\mathrm{P_1}\mathrm{exp}\left\{-\frac{1}{2}\left(\frac{x-\mathrm{P_2}}{\mathrm{P_3}}\right)^2\right\} \nonumber \\
+\mathrm{P_4}\mathrm{exp}\left\{-\frac{1}{2}\left(\frac{x-\mathrm{P_5}}{1.3 {\rm arcmin}}\right)^2\right\} \nonumber \\
+2.3\mathrm{P_4}\mathrm{exp}\left\{-\frac{1}{2}\left(\frac{x-\mathrm{P_5}}{0.7 {\rm arcmin}}\right)^2\right\} .
\end{eqnarray}
Here, $\mathrm{P_0}$, $\mathrm{P_1}$, and $\mathrm{P_4}$ are the intensity of
background, diffuse, and point source (the pulsar), respectively, 
in the unit of $10^{-5}$counts~s$^{-1}$,
$\mathrm{P_3}$ is the extent of the diffuse emission
in the unit of arcmin,
and $\mathrm{P_2}$ and $\mathrm{P_5}$ is the offset of
diffuse and point source from the arbitral origin 
in the unit of arcmin.
In this method, we do not have to consider
whether the remaining hard-tail emission in the background region
is real low-surface brightness component or just due to stray light,
since it does not change the width of the Gaussian fitting.

The fitting results are summarized in table \ref{tab:linirprofile}.
The fit gave the size of the diffuse emission ($P_3$) of
\timeform{4'.1} $\pm$ \timeform{1'.4} (region A) and
\timeform{4'.5} $\pm$ \timeform{0'.4} (region B), respectively.
The center position of the diffuse emission was
away from that of the pulsar by \timeform{1'.7} $\pm$ \timeform{0'.6}
(region A)
and \timeform{1'.3} $\pm$ \timeform{0'.3} (region B), respectively.
The center of the diffuse emission is shown as a diamond in figure~\ref{fig:xisimage}.

\begin{longtable}{*{8}{c}}
\caption{Fitting results of the one-dimensional profiles}
\label{tab:linirprofile}
\hline
\hline
Region & P$_0$ & P$_1$ & P$_2$ & P$_3$ & P$_4$ & P$_5$ & $\chi^2$(d.o.f.)   \\
 & [$10^{-5}$counts~s$^{-1}$] & [$10^{-5}$counts~s$^{-1}$] & [arcmin] & [arcmin] & [$10^{-5}$counts~s$^{-1}$] & [arcmin] & \\ 
\hline
A & $2.0\pm1.1$ & $3.2\pm1.0$ & $8.0\pm0.3$ & $4.1\pm1.4$ & $0.26\pm0.16$ & $9.7\pm0.5$ &10.5(24) \\
B & $1.5\pm0.1$ & $3.1\pm0.2$ & $13.2\pm0.2$ & $4.5\pm0.4$ & $0.60\pm0.15$ & $11.9\pm0.2$ &26.2(43) \\
\hline
\endlastfoot
\end{longtable}

\subsection{Energy spectrum}
\subsubsection{Source spectrum}
\label{sec:spec_total}
\begin{figure}
  \begin{center}
    \includegraphics[width=80mm]{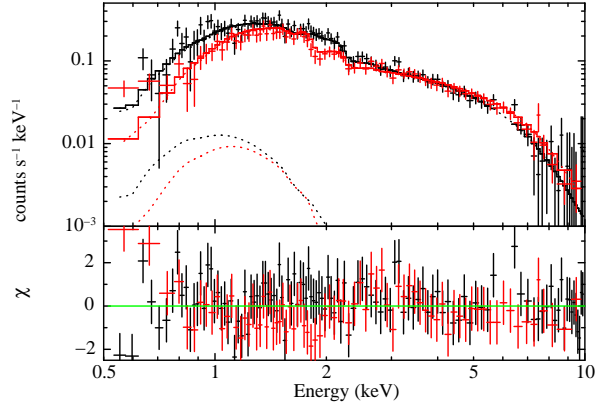}
  \end{center}
  \caption{%
Energy spectrum of the source region.
The best-fit model consisting of a power-law (solid) and
NSA (dotted) is also shown.
Black and red crosses show the BI and FI spectra, respectively.
Vertical error bars of each data point represent a 1$\sigma$ error.
}
\label{fig:sourcespec}
\end{figure}

In order to study the X-ray properties of the extended PWN,
we need to exclude the emission from the central pulsar.
However, it makes difficult to study the central part of PWN,
due to the lack of spatial resolution of Suzaku.
We thus first studied
the energy spectrum of the total emission, sum of the pulsar
and the diffuse emission, with the Suzaku data.
We extracted the source spectrum from a \timeform{10'}-radius circle
depicted with the yellow circle in figure~\ref{fig:xisimage}.
As well as the image analysis, we removed the data of the point-like sources
with a \timeform{1'}-radius circle.
We subtracted the background spectrum consisting of the NXB
and the sky background emission discussed in \S\ref{sec:bgspectrum}.
The NXB spectra were calculated using \texttt{xisnxbgen}.
We compared the NXB and source spectra and found that
the count rate of the calculated NXB is $\sim 83\%$ of that of the observed one
above 10~keV, where NXB is the dominant component.
It is considered the uncertainty of the NXB reproduction procedure
(Mori et al., private communication),
we adjusted the simulated NXB to fit the observed spectra
above 10~keV.
The sky background spectra were simulated using \texttt{xspec}
by assuming the input energy spectrum extracted from the background region
as shown in \S\ref{sec:bgspectrum}.
The simulation was necessary to correctly incorporate the difference of the
extraction regions and the energy dependent vignetting.
This is the same method applied in \citet{2013PASJ...65...61F}.
The exposure of the simulation was set to the same value as
that of the Suzaku observations.
The background subtracted spectrum is shown in figure~\ref{fig:sourcespec}.

The spectrum was fitted with
an absorbed power-law model in the 0.5$-$10.0~keV band.
The source region contains the pulsar.
According to the spectral analysis of the Chandra observation
\citep{2012ApJ...744...81C},
an emission from the pulsar itself was described with a thermal emission from the
hydrogen atmosphere of a magnetized neutron star.
Therefore, we added a NSA model to the above model.
We adopted the NSA's parameters from the Chandra observation,
$T_{\mathrm{eff}}=0.96$~MK, $M_{\mathrm{ns}}=1.4\MO$,
$R_{\mathrm{ns}}=10$~km, $B=$~10$^{13}$~G, and $D=$2.4~kpc.
The fit was acceptable with a reduced $\chi^2$ of 221.6/253.
The best-fit parameters are summarized in table \ref{tab:sourcespecresults}.  
Note that the emission is more than one order of magnitude
brighter than the power-law component in the background region,
thus we can see that the source region cover
the most part of the PWN emission.

\begin{longtable}{*{4}{c}}
\caption{Best-fit parameters of the spectrum in the source region.}
\label{tab:sourcespecresults}
\hline
\hline
$N_{\mathrm{H}}$\footnotemark[a] & $\Gamma$\footnotemark[b] & $F_{\mathrm{unabs}}$\footnotemark[c] &$\chi^2_{\nu}$(d.o.f.)   \\
\hline
$0.35^{+0.05}_{-0.04}$ & $1.70^{+0.07}_{-0.07}$ & $7.9_{-0.3}^{+0.3}$ & 0.88(253)  \\
\hline
\multicolumn{4}{l}{\footnotemark[a] Column density [$10^{22}$cm$^{-2}$] } \\
\multicolumn{4}{l}{\footnotemark[b] Photon index} \\
\multicolumn{4}{l}{\footnotemark[c] Unabsorbed flux (2.0$-$10.0 keV) [$10^{-12}$erg s$^{-1}$ cm$^{-2}$]} \\
\endlastfoot
\end{longtable}

\subsubsection{Spatial variations of the spectrum}
\begin{figure}
  \begin{center}
    \includegraphics[width=80mm]{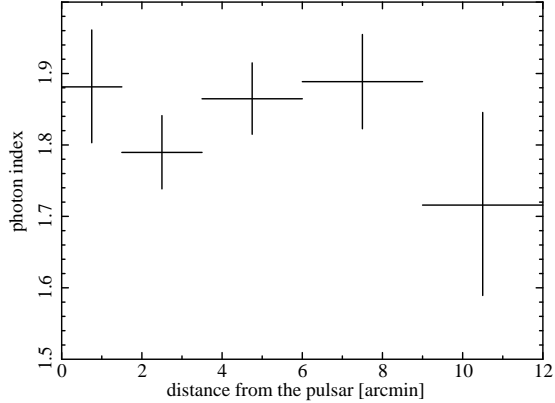}
  \end{center}
  \caption{Spatial variation of the photon indices from the pulsar PSR J1357$-$6429. 
  No significant spatial variation was obtained.}
  \label{fig:photonindex_dependance}
\end{figure}

In order to investigate the spatial variations of the spectral shape,
we divided the spectral extraction region into five annular ones,
i.e., "region~1" for the radius range of  \timeform{0'}$-$\timeform{1'.5},
"region~2" for \timeform{1'.5}$-$\timeform{3'.5},
"region~3" for \timeform{3'.5}$-$\timeform{6'},
"region~4" for \timeform{6'}$-$\timeform{9'},
and "region~5" for \timeform{9'}$-$\timeform{12'}.  
As well as the analysis of the source spectrum,
the point-like sources were removed and
the NXB and the sky background spectra were subtracted.
We fitted each spectrum with an absorbed power-law model
in the 0.5$-$10~keV band.
In the fitting, we fixed the hydrogen column density at
the average value, $N_{\mathrm{H}}=3.5\times 10^{21}$~cm$^{-2}$,
determined by the  spectrum analysis of entire emission
in \S\ref{sec:spec_total}.
to reduce the uncertainty due to limited statistics.
For the region~1, we also added a fixed NSA component
to reproduce the pulsar emission.
The best-fit photon indices  are shown 
in Figure~\ref{fig:photonindex_dependance} as a function of the distance 
from the pulsar.
We can see that no significant spatial variation
was detected in the photon indices.

\section{Discussion}
\label{sec:discussion}

%\subsection{Distance to the extended emission}
%
%We found the absorption column of the extended emission 
%to be $3.5\times 10^{21}$~cm$^{-2}$
%as listed in Table~\ref{tab:sourcespecresults}.
%If we assume that a mean density of the interstellar medium (ISM) is
%$\sim$0.5--1.0~cm$^{-3}$,
%the distance to the extended emission is estimated to be roughly 1.1--2.3~kpc.
%However, because the ISM density is on average higher
%near the Galactic plane and
%decrease rapidly away from the plane, the estimated distance may be 
%regarded as a lower limit.
%On the other hand,
%if the source has an intrinsic absorption, the distance estimation 
%would become smaller.
%In fact, the galactic absorption toward the source is
%estimated to be $6.0\times 10^{20}$~cm$^{-2}$ \citep{2005A&A...440..775K},
%which is smaller than the observed value.
%Considering such ambiguity, we have to say that the absorption column
%we obtained may still be consistent to the distance to the pulsar, 2.4~kpc
%\citep{2002astro.ph..7156C}.
%In what follows, we assume that the source of the extended emission is located
%at the same distance to PSR~J1357$-$6429.

%\subsection{The extended power-law emission}

We have discovered an extended hard emission
around the pulsar PSR~J1357$-$6429.
Because the emission has a power-law spectrum and is
roughly centered at the pulsar, we consider that the emission 
is originated from the PWN associated to the pulsar.
The emission size is $\sim$4~arcmin, or $\sim$3~pc at 2.4~kpc
\citep{2002astro.ph..7156C},
which suggests that
the PWN is one of the most extended samples 
among the young PWNe with the age of less than 10~kyr
\citep{2010ApJ...719L.116B}.

\citet{2010ApJ...719L.116B} discussed that PWNe evolves up to 
the age of $\sim$100~kyr continuously expanding in size.
The evolution may reflect the decay of magnetic field
in PWNe and electrons can escape from the system
before loosing their energy.
The PWN of PSR~J1357$-$6429 follows the relation
between
the age of the system (characteristic age of the pulsar) 
and the extent of the PWN
suggested in \citet{2010ApJ...719L.116B}.

Relative positions of the center of the PWNe in different wavebands and 
the pulsar may be an important clue to understand the
evolution of the PWNe.
The center of the extended X-ray emission is found to be largely separated from the
center of HESS J1356$-$645 in the TeV band, which coincide the radio center
of the PWN \citep{1995MNRAS.277...36D}.
The X-ray and TeV centers of the PWN are both offset from the pulsar, 
but the X-ray center is much closer to the pulsar than the TeV center (see Figure~\ref{fig:xisimage}).
The direction of the pulsar offset from the TeV center is similar to that of 
the proper motion of the pulsar \citep{2012A&A...540A..28D}.
These facts may imply that the pulsar
%If we trace the movement of the pulsar backwards, it may
have traveled from 
the center of the TeV emission through the vicinity of X-ray center toward the current position.
%The X-ray center is slightly offset from the line connecting the pulsar and 
%the TeV center.
%This may be due to the systematics in determining the X-ray center; 
%a simple gaussian model was used to fit the profile, while the real 
%morphology of the extended X-ray emission could be more complex.
Such configuration of the PWN emission in each band
and the pulsar may conform to
the so-called relic PWN scenario
\citep{2001ApJ...563..806B,2006ApJ...636..777A,2007ApJ...670..655K}.
According to the scenario, the pulsar was born at the center of the TeV (and radio) 
emission with a supernova explosion 7.3~kyr ago,
and reached the current position running toward the northeastern direction
with the fast proper motion of $\sim$2000~km~s$^{-1}$
\citep{2012A&A...540A..28D}.
The pulsar accelerated electrons during the movement, which diffused out forming a PWN\@.
When we consider the center position of the PWN,
it is important to estimate the cooling times
of energetic electrons.
High energy electrons lose their energy quickly via the synchrotron emission,
while less energetic electrons lose energy more slowly.
Typical energy of X-ray emitting electrons, whose emission mechanism is via synchrotron 
process, may be $\sim$80~TeV for the assumed magnetic field of 10~$\mu$G 
(see eg.\ \cite{2010ApJ...719L.116B}).
On the other hand, the TeV gamma-ray emission may be produced by
inverse Compton scattering of cosmic microwave background radiation 
by $\sim$10~TeV electrons.
As a result, TeV emission sustains longer at the pulsar birthplace,
whereas the X-ray emission is confined only nearby the present pulsar position.
This means the X-ray emitting electrons are those accelerated recently.

The center of X-ray emission is offset from the pulsar by $\sim$2~arcmin.
Because the X-ray center is calculated assuming a simple gaussian model and
the direction of the offset is different from that of the proper motion of the pulsar,
we consider that amount of the offset should be regarded just as a representative value.
In spite of this ambiguity,
we can still argue that the offset is much smaller than that of
the TeV center ($\sim$7~arcmin).
The X-ray offset is about one-third of the TeV offset.
Assuming that the electrons emitting X-rays and TeV gamma-rays
are accelerated around the pulsar and diffuse out in the same speed,
and that electrons accelerated when the pulsar was very young still emit TeV
gamma-rays, 7.3~kyrs ago (pulsar age),
we can estimate the typical age of electrons emitting X-rays
to be one-third of the pulsar age, very roughly, $\sim$2~kyr ago.
Note that the pulsar spin-down age can differ by a factor 2--3
from true age, and the quantitative estimation of such numbers is
very difficult.

This interpretation is also supported from the difference of the sizes
of the emission regions.
The extent of X-rays is $\sim$4~arcmin, which is again about one-third of
the TeV extent of 0.2~degree again.
This difference is naturally explained if the X-ray emitting electrons
were accelerated recently.

We detected no significant spatial variations in the energy spectrum
of the extended X-ray emission (Figure~\ref{fig:photonindex_dependance}).
This result 
cannot be interpreted in a simple way
with the relic PWN scenario.
The constant photon index means that the X-ray emitting electrons
do not lose significant energy via synchrotron cooling
during the diffusion from the pulsar,
which conflicts with the offset of X-ray and TeV PWNe.
This apparent conflict may be resolved if we introduce 
a sudden increase of acceleration
efficiency when 
the reverse shock hit the PWN \citep{2009ApJ...703.2051G}.
Another possibility is that the acceleration efficiency became higher
when the pulsar and dense interstellar matter collided to produce
a strong bow-shock due to the fast proper motion of the pulsar.
%If we follow these scenarios, the diffusion time scale of
%electrons becomes not 7.3~kyr but $\sim$2~kyr.
%Then, the average diffusion velocity would be very fast,
%$\sim 2\times 10^8$~cm~s$^{-1}$,
%which is much higher than that estimated in \citet{2010ApJ...719L.116B}.
%It is not clear at present whether an unknown mechanism exists to allow very
%fast diffusion of energetic electrons,
%or the relic PWN scenario does not hold true.
%Further multi-wavelength studies of PWNe are needed.
We need the magnetic field strength, configuration, and turbulence
in order to estimate the diffusion time scale,
which are beyond the scope of this paper,
since all of these parameters are time-dependent \citep{2010ApJ...719L.116B}.
Further multi-wavelength studies of PWNe are needed.

\bigskip

We thank the anonymous referee for fruitful comments.
M.I. and A.B. thank Makoto Sawada
for his help on the Suzaku data analysis.
We also thank Koji Mori for the useful comments on the NXB spectra.
This work was supported in part
by Grant-in-Aid for Scientific Research of the Japanese Ministry of
Education, Culture, Sports, Science and Technology,
No.~22684012 (A.~B.).

%%%
% See the manual for the detail.
%%%

\end{document}